# Investigating the gaze control ability of VALORANT players using a Python-based tool

Inhyeok Jeong, *The University of Tokyo,* Takuma Nobuto, *Keio University*, Naotsugu Kaneko, *The University of Tokyo*, Takaaki Kato, *Keio University*, and Kimitaka Nakazawa, *The University of Tokyo*

*Abstract*—The current study investigated the gaze movements of FPS gamers in actual game environments. We developed a low-cost analysis tool using Python to identify gaze movements in real-world gaming environments. In Experiment 1, 11 middle-skilled and ten high-skilled FPS gamers performed a task under the experimental condition. Gaze position, reaction time, and accuracy were calculated during the task. Reaction time exhibited a significant positive correlation with task accuracy, suggesting that speed and accuracy were associated with higher game performance. The middle-skilled gamers had a significantly wider horizontal gaze distribution than the high-skilled gamers, and gaze distribution and reaction time showed a negative correlation. These results suggested that high-skilled players utilize peripheral vision during gameplay. In Experiment 2, 15 middle-skilled and 12 high-skilled FPS gamers performed an actual FPS game match. The gaze distribution, kill/death/assist ratio (KDA), and percentage of gaze on game information were calculated. In experiment 2, gaze locations in less important areas were positively correlated with KDA. Thus, performance was determined by the important areas where the gaze was focused rather than by the coordination of gaze position alone. Therefore, a broader range of environments is necessary to comprehend the superior performance of FPS gamers.

This work was supported Tateisi Science and Technology Foundation, Shimogyo, Kyoto [grant number 2237004]. There was no Declarations of Interest.

*Corresponding author: Kimitaka Nakazawa*
Graduate School of Arts and Sciences, The University of Tokyo
3-8-1, Komaba, Meguro-ku, Tokyo 153-8902, Japan
Tel.: +81-3-5454-6014; Fax: +81-3-5454-6014
E-mail: nakazawa@idaten.c.u-tokyo.ac.jp

First author: Inhyeok Jeong is with the Graduate School of Arts and Sciences, The University of Tokyo, Tokyo, Japan (email: hofwillson@g.ecc.u-tokyo.ac.jp).
Second author: Takuma Nobuto is now with the Faculty of Environment and Information Studies, Keio University, Kanagawa, Japan and BetterClick Co., Ltd, Tokyo, Japan (email: takuma130625@keio.sfc.jp).
Third author: Naotsugu Kaneko is now with the Graduate School of Arts and Sciences, The University of Tokyo (email: kaneko@idaten.c.u-tokyo.ac.jp).
Fourth author: Takaaki Kato is now with the Faculty of Environment and Information Studies, Keio University, Kanagawa, Japan (email: tiger@sfc.keio.ac.jp).
Fifth author: Kimitaka Nakazawa is now with the Graduate School of Arts and Sciences, The University of Tokyo (email: nakazawa@idaten.c.u-tokyo.ac.jp).

There were no supplemental materials included.

## I. INTRODUCTION

FIRST-person shooter (FPS) is a genre of esports in which players fight enemies with gun-based weapons from a first-person perspective. As FPS games gain popularity, an increasing number of studies have been conducted to identify the root of the superior performance of high-skilled FPS game players [1]. In FPS games, fast reaction time, kill-death-assist count, and game score were used to evaluate the performance level of esports players [2]. Previous studies have examined the cognitive functions of high-skilled FPS game players, such as faster reaction time, information processing skills, and visual sensitivity [3, 4, 5]. Although studies have shown the outstanding cognitive function of high-skilled FPS game players, the gaze control ability related to superior performance is still unclear. Gaze control ability is the cognitive and motor process of directing visual attention to a specific object or location [6].

Examining gaze movement helps understand the superior performance of general athletes [7]. A systematic review has shown a strong relationship between gaze movements and sports performance [8]. For example, increasing the fixation duration of the gaze position can increase the success rate of free throws in basketball [9]. In addition to general sports, esports studies examining gaze movements have identified the source of the superior performance of esports players. For example, skilled esports players have a wider gaze distribution than low-skilled players in a controlled environment [10]. In FPS games, high-skilled FPS game players fixate their gaze position on the task more than low-skilled FPS game players during the precision task [11]. Moreover, high-skilled FPS game players can respond more rapidly to visual stimuli than non-FPS gamers [12, 13]. In sum, high-skilled FPS game players have superior gaze control abilities than non and low-skilled FPS game players.

However, previous research has two common limitations. First, previous research did not examine the gaze control ability of FPS game players in actual FPS game environments. Previous studies examined FPS game players' performance using only simple tasks that could be used in laboratory environments [11, 12]. Since the actual FPS game and the experimental environments have different characteristics [14] and the actual FPS game requires complex strategies, it remains unclear whether superior gaze control abilities in the experimental environments extend to actual game environments. Previous research has pointed out that examining performance levels in real-world sports environments can help improve the reliability and validity of data obtained in controlled environments [15]. Second, no studies have been conducted on the differences between middle-skilled and high-level FPS players. Previous studies only compared the superior gaze control ability and visual attention skills in low- and high-skilled FPS game players [11, 13]. However, in actual FPS games, there are more matches between skilled FPS players than between low- and high-skilled FPS game players. Therefore, it is important to compare the gaze control ability between middle- and high-skilled FPS players, and it can help understand the superior performance of skilled FPS game players.

Thus, the current study aimed to investigate the characteristics of the gaze movements of skilled FPS game players in actual and controlled environments. We used one of the most famous FPS games, VALORANT (Riot Games, California, USA), as an actual FPS environment. VALORANT is an FPS game that teams up to 5 players to fight against an opponent. In addition, VALORANT requires the use of appropriate skills and good teamwork to win the match. Furthermore, we developed a low-cost gaze movement analyzing tool called "get_eyedata" using Python (an open-source programming language) and combined it with an inexpensive eye tracker (Tobii 4C) to conduct experiments in an environment similar to an actual FPS game (i.e., a controlled environment). The combined tool can calculate gaze coordinate values and regions of interest (ROIs) to analyze useful information from FPS games. The findings and developed tool in the current study will help reveal the gaze control abilities of FPS game players under the actual FPS game environment and for developing future FPS game-specific training methods.

## II. METHOD

### A. Experimental preparation

The experiment was conducted in GAKU Bootcamp Academy 2023, held in Toyo, Japan (March 23, 2023). The experiment was divided into two parts (Experiment 1 and Experiment 2). Figure 1 shows the experimental preparation and overall flow of the experiment.

Before starting the experiment, the experimental environment was prepared as follows: First, the monitor (I-O DATA, 24 inches, 144 Hz) and Tobii 4C eye tracker (Tobii Technology Inc., VA, USA) were set for the experiment (sampling rate: 90 Hz; 100kB/s USB data transfer rate). The Tobii 4C has been used in previous studies to measure and evaluate gaze movements [16,17,18]. Participants could use a personal mouse and keyboard. Second, Tobii Ghost and Experience (Tobii Technology Inc., VA, USA) and Open Broadcaster Software (OBS Studio) were used to record gaze movement data. Tobii Ghost and Experience were used to represent gaze movement in the recording video and for calibration, respectively. OBS Studio was used to record the gaze position and gaming scenes. All participants used Tobii Ghost, Experienced, and OBS Studio to record their gaze movement via the video file (Table. 1). Third, the screen recording environment was set to 1920x2160 pixels. The upper part of the monitor was set for the actual game scene to appear, and the lower part was set to show gaze movement (Figure 1B). Participants could only see the upper part of the monitor. Fourth, when the experimental environment was prepared and participants were ready to start the game, participants pressed the record button and played the game. All video recordings were created as .mkv files. The overall flow of the experiment is presented in Figure 1C.

Finally, after the experiment, gaze movement data were exported by "get_eyedata" (described in *The analysis tool "get_eyedata"* section). The analysis tool "get_eyedata" calculated the coordinates of gaze position (vertical and horizontal values) and the percentage of gaze movements that stopped on important in-game information, called regions of interest (ROIs). All exported data were generated by Excel .csv file. After the experiment, we performed a power analysis to determine the power of the sample size (G*Power version 3.1.9). The power analysis was conducted using the reaction time of Experiment 1 (Cohen's d: 1.04; α level: 0.05; a sample size of the High Skill group: 10; a sample size of the Middle Skill group: 11). Effect size was calculated according to Cohen's method [19]. According to the result of the power analysis, the actual power was 0.624 (noncentrality parameter δ: 2.39; Crital t: 2.09; Df: 19).

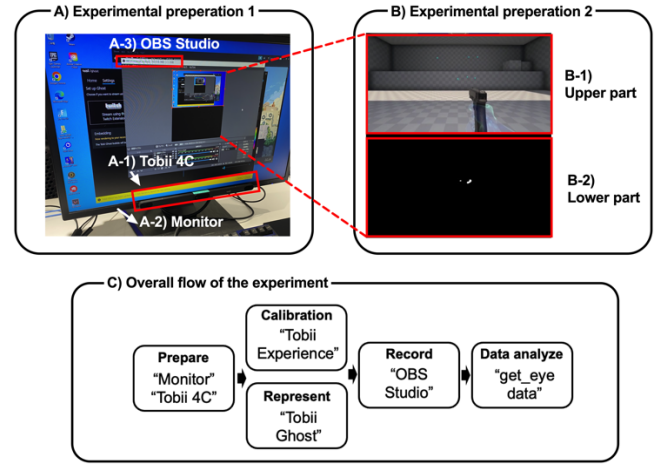

**Figure 1.** Experimental preparation and overall flow. A) OBS Studio, Tobii 4C, and Monitor. B) Features of the recording environment of OBS Studio. C) Overall flow of the experiment.

TABLE I
ROLE OF EACH DEVICE

| Device | Role |
|---|---|
| Tobii 4C | Used for measuring the gaze movement |
| Tobii Ghost | Used for representing the gaze position in the recording video |
| Tobii Experience | Used for calibrating the gaze movement |
| OBS Studio | Used for recording the represented gaze movements via the video file |
| get_eyedata | Calculate coordinates and ROI of gaze movements recorded in video files |

### B. The analysis tool "get_eyedata"

We developed a low-cost gaze movement analyzing tool called "get_eyedata" using Python. The gaze movement analysis tool "get_eyedata" can be downloaded from the following link:

https://github.com/ikrfun/get_eyedata

Specific installation methods and the user manual of "get_eyedata" can be checked in the following link:
https://github.com/ikrfun/get_eyedata/blob/main/README.md

Before analyzing the gaze movement, it is necessary to prepare the gaze movement recording file. Instructions for preparing the data can be accessed at the following link:

https://github.com/ikrfun/get_eyedata.wiki.git

The analysis tool "get_eyedata" is released under the MIT license and requires Python version 3.10.6 to use. Anyone can use "get_eyedata" freely. In this tool, the gaze position (white dot) was tracked through the tracking point (green dot), and the location was expressed as a number between 1920 (X-axis) and 1090 (Y-axis). Moreover, the ROI was categorized into five different areas: center, mini-map, other, information 1, and information 2 (Figure 2A). Information about team members and remaining time can be checked in information 1. In information 2, the number of left bullets, health points, and skills were contained.

The analysis tool "get_eyedata" extracted the frame from the gaze movement video recorded in the experiment. At the end of the extraction, the gaze movement data included in each frame were output in Excel. csv file; the Excel file was saved as "output.csv" (Figure 2B-3). In the exported Excel file, the coordinates of gaze position, ROI information, and frame ID were recorded (Figure 2C).

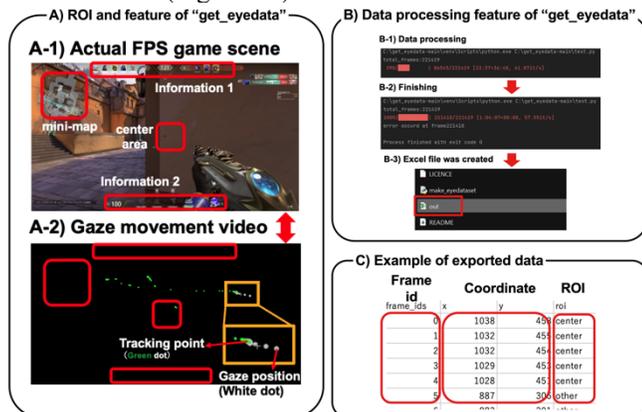

**Figure 2.** ROI and information about "get_eyedata". A) Each red box indicates the ROI: mini-map, center area, information 1, and information 2. B) Each figure indicates the data processing feature of "get_eyedata". C) Example of an exported Excel data file. In these data, the frame id, coordinate of gaze position, and ROI were calculated.

*C. Participants*

All participants were college students (1st to 4th year) who did not have any gaming disorders and had experience playing VALORANT. All participants had normal or corrected-to-normal vision. VALORANT is an FPS game in which a team of five players fights against an opponent team. During the experiment, we instructed participants to keep the distance between their heads and the monitor as same as possible (100 cm).

In Experiment 1, eleven middle-skilled FPS game players (the Middle Skill group) and ten high-skilled FPS game players (the High Skill group) were recruited. One Silver rank player, one Gold rank player, five Platinum rank players, and four Diamond rank players belonged to the Middle Skill group. Four Ascendant rank players, five Immortal rank players, and one Radiant rank player belonged to the High Skill group. In Experiment 2, fifteen Middle Skilled VALORANT game players (two Silver rank players; two Gold rank players; six Platinum rank players; five Diamond rank players) belonged to the Middle Skill group, and twelve high-skilled VALORANT game players (six Ascendant rank players; five Immortal rank players; one Radiant rank player) belonged to the High Skill group. Participants played an average of 1.17 (SD ± 0.5) genre games, including FPS games. In the average experience years, the High Skill group had 4.3 (SD ± 4.6) years of playing FPS games, and the Middle Skill groups had 3.6 (SD ± 2.1) years of experience playing FPS games. All participants used their mouse and keyboard. Participants used five different brands of gaming mice (Logitech International S.A, Lausanne, Switzerland; Razer, California, USA; HyperX, California, USA; SteelSeries, Frederiksberg, Denmark; BenQ Corp. Neihu, Taiwan). According to the manufacturer user manual, all mice had a low response latency (1 millisecond). Therefore, response latency did not significantly affect reaction time during the experiment. To avoid the placebo effect, participants did not know which group they belonged to (Middle Skill or High Skill). Before the experiment started, all participants were given the contents and concepts of the experiment and asked to sign a written consent form. All experimental procedures were approved by the local ethics committee of the University of Tokyo (approval number: 912).

*D. The procedure of Experiment 1*

In Experiment 1, the participants played the task provided by a training program for FPS game players called "AimLab" (State Space Inc., New York, USA). AimLab is a tool designed to help FPS game players improve their reaction time and game performance (e.g., increase the reaction time fast and react accurately). Participants only used mice for Experiment 1. In the experiment, participants were divided into skill levels (the High Skill and Middle Skill groups) and instructed to hit the appearing targets as quickly and accurately as possible using the left mouse click. The task was performed for 1 min in five trials (a total of 5 min). The standard deviation of gaze movement (SD of X position and SD of Y position), and the distance from the center of the monitor to the average gaze position were exported by "get_eyedata". Performance levels (reaction time and accuracy) were automatically calculated within the AimLab program at the end of the task. The features of the task in Experiment 1 can be checked in Figure 3.

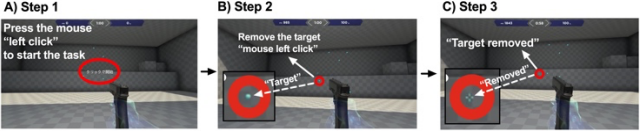

**Figure 3.** The overall flow of the task. Each figure shows the flow of the task. A) The preparation phase of the task. B) Features of the target removal process. C) Feature of the removed target.

*E. The procedure of Experiment 2*

Experiment 2 was conducted after Experiment 1. In Experiment 2, participants played a 5-on-5 practice match in VALORANT. The practice match consisted of 17-26 trials (a total of 529 trials) for each participant in the High Skill and Middle Skill groups. The number of trials was not constant because the number of match point trials changed based on the result of the practice match. Each trial in the practice match was performed for 1 minute and 40 seconds. The opponent players each participant played against were randomly assigned. Before starting the experiment, we instructed the participants to win as many matches as possible. Participants could freely select the characters and weapons they wished to use. During Experiment 2, participants used a mouse left-click to shoot the weapons. A mouse right-click was used to zoom in on the enemy characters. Keyboard w, a, s, and d keys were used to move the characters of the participants. Moreover, participants were allowed to set up their own character's skills on the keyboard. In Experiment 2, ROI and gaze distribution (vertical and horizontal gaze distribution) were exported after the practice match through "get_eyedata". In addition, the performance level of Experiment 2 was evaluated by the Kill/Death/Assist ratio (KDA) when each participant played the practice match. KDA was automatically calculated at the end of a practice match in the VALORANT gaming system. KDA is a way to evaluate the performance level based on the number of knockouts of the opponent, the number of knockouts by an opponent, and the number of contributions to the knockout of an opponent. KDA was calculated by the following equation (1):

$$\text{(Number of knocked out the opponent + Number of contributed to the Knockout of and opponent)}/\text{(Number of knocked out by and opponent)} \quad (1)$$

*F. Statistical analysis*

All statistical analyses were performed using R Studio version 2022.02.1+461 (R Studio, Boston, MA, USA). In Experiment 1, performance data (accuracy and reaction time), gaze distribution, and distance from the center of the monitor to the average gaze position were calculated. First, we checked for normality and homogeneity using the Shapiro-Wilk test and Levene's test, respectively. These tests showed that the performance data (accuracy and reaction time) followed normality and homogeneity. Thus, an unpaired t test was performed to compare the performance level between the High Skill and Middle Skill groups. The correlation coefficient was calculated between performance data (reaction time and accuracy) by using the Pearson correlation (Pearson's *r*). The gaze distribution and distance from the center of the monitor to the average gaze position, which did not follow normality and homogeneity, were compared between the High Skill and Middle Skill groups by the Wilcoxon rank sum test. The correlation coefficient between the performance data and gaze movement data (i.e., gaze distribution and distance from the center of the monitor to the average gaze position) was calculated by using Spearman's rank correlation (Spearman's ρ).

In Experiment 2, the Shapiro–Wilk test and Levene's test showed that all data did not follow normality and homogeneity. Thus, at the performance level, the KDA of Middle Skill and High Skill were compared by using the Wilcoxon rank sum test. In the gaze movement data, ROI and gaze distribution between the High Skill and Middle Skill groups were calculated by the Wilcoxon rank sum test. Finally, Spearman's rank correlation coefficient (Spearman's ρ) was calculated to examine the correlation between KDA and gaze information (ROI and gaze distribution). All significance levels were set at $p < .05$.

### III. RESULT

*A. Experiment 1*

Figure 4 shows the differences between the High Skill and Middle Skill groups and the correlation between reaction time and accuracy. Figure 4A and 4B represent the data of each trial. The unpaired t test found that the reaction time of the High Skill group was significantly faster than that of the Middle Skill group (t = 2.21, p =.03, Figure 4C). However, no significant difference was found in accuracy (t = -.94, p =.35, Figure 4D). The Pearson correlation coefficient revealed a strong negative correlation between reaction time and accuracy (r = -0.581, t = -3.11, p <.001, Figure 4E).

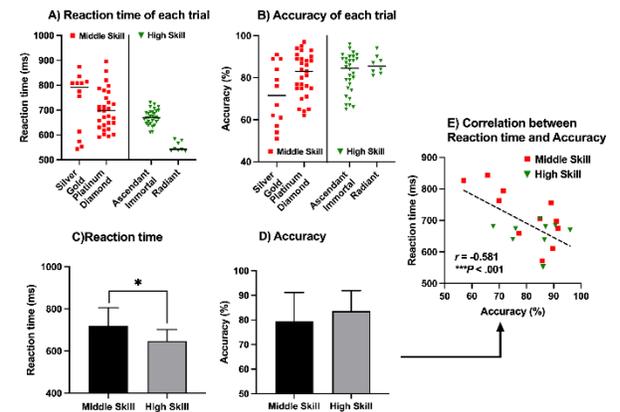

**Figure 4.** Results of the statistical analysis at the performance level. A and B) Each dot shows the data of each trial. The line in the plot indicates the mean value of the data. C and D) Each box indicates the average reaction time and accuracy. Whisker represents the standard deviation of the data. The significance level was * *p* <.05. E) The dashed line represents the linear regression between reaction time and accuracy. The significance level was *** $p <.001$.

Figure 5 shows the gaze distribution and the correlation between gaze distribution and performance level. Figure 5A and 5E indicates the horizontal and vertical gaze distribution of each trial. In the gaze distribution, the Middle Skill group showed a significantly wider horizontal gaze distribution than the High Skill group (Wilcoxon rank sum test; p =.002, Figure 5B). Moreover, reaction time positively correlated with horizontal gaze distribution (Spearman's ρ = 0.573, p =.008, Figure 5C). There was no significant correlation between horizontal gaze distribution and accuracy (Spearman's ρ = -0.315, p =.16, Figure 5D). Vertical gaze distribution did not significantly differ between the High Skill and Middle Skill groups (Wilcoxon rank sum test; p =.051, Figure 5F). Additionally, vertical gaze distribution and reaction time had a significant positive correlation (Spearman's ρ = 0.499, p =.02, Figure 5G). However, there was no significant correlation between vertical gaze distribution and accuracy (Spearman's ρ = -0.143, p =.53, Figure 5H).

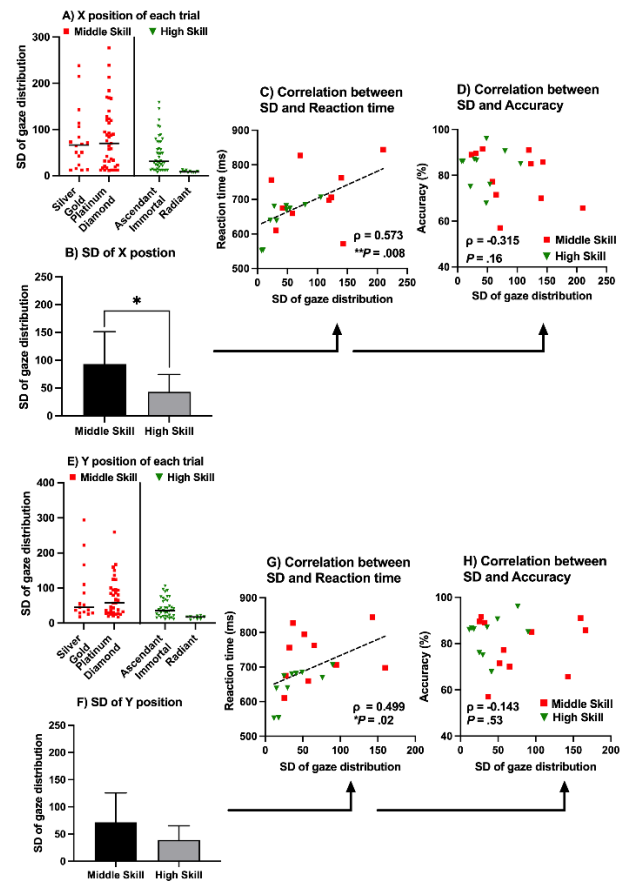

**Figure 5.** Gaze distribution data and correlation between performance level and gaze distribution. A and E) Each dot shows the data of each trial. The line in the plot indicates the mean value of the data. B and F) Each bar in the graph shows the average SD of gaze distribution. Whisker indicates the standard deviation of the data. C, D, G, and H) Dashed lines represent the linear regression of each data point. The significance levels were * p <.05 and ** $p <.01$.

In the distance from the center of the monitor to the average gaze position, no significant difference was found between the High Skill and Middle Skill groups (Wilcoxon rank sum test, p =.28, Figure 6B). However, reaction time and distance between the center of the monitor and the average location of gaze position were significantly correlated (Spearman's ρ = 0.450, p =.04, Figure 6C).

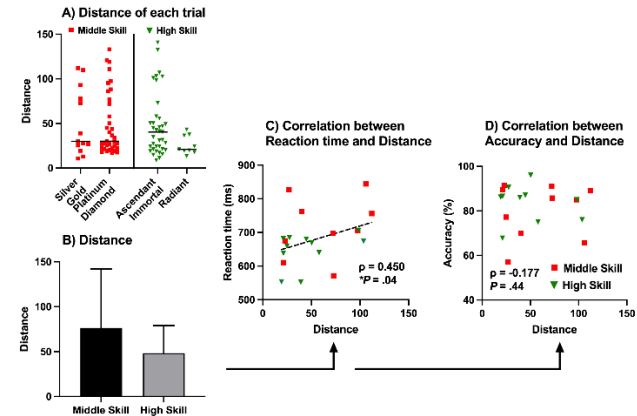

**Figure 6.** Distance between the center of the monitor and the average location of gaze position. A) Each dot shows the data of each trial. The line in the plot indicates the mean value of the data. B) The bar in the graph shows the average of the data. Whisker indicates the standard deviation. B) A dashed line represents the linear regression of the data. The significance level was * p <.05.

*B. Experiment 2*

Figure 7 represents the KDA of each group and the correlation between KDA and gaze movement data. There was no significant difference between the Middle Skill and High Skill groups in KDA (p =.53, Figure 7A). Moreover, there was no significant correlation between KDA and each ROI (Center: Spearman's ρ = 0.049, p =.81; Mini-map: Spearman's ρ = 0.111, p =.59; Other: Spearman's ρ = -0.035, p =.86; Information 2: Spearman's ρ = -0.013, p =.94). However, Information 1 and KDA had a significantly negative correlation (Spearman's ρ = -0.608, p =.002, Figure 7E). In the correlation between KDA and gaze distribution, no significant difference was found (SD of X position: Spearman's ρ = -0.121, p =.56; SD of Y position: Spearman's ρ = 0.068, p =.74).

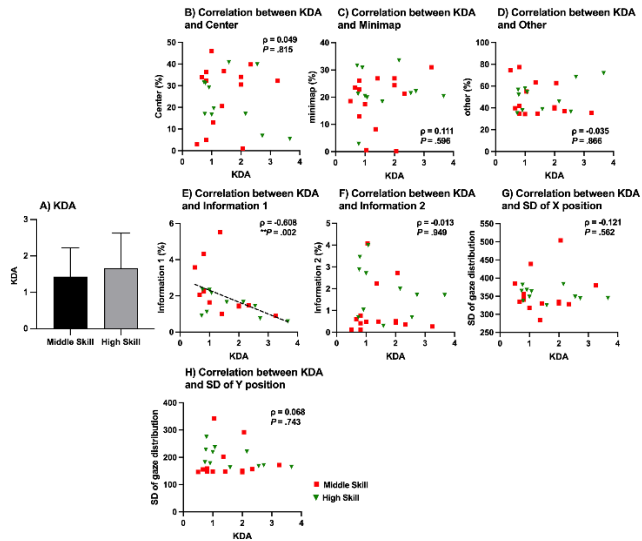

**Figure 7.** Statistical analysis of KDA and gaze movement. A) Each bar indicates the average KDA. Whisker shows the standard deviation of KDA. B, C, D, E, F, G, and H) A dashed line represents the linear regression of the data. The significance level was ** $p$ <.01.

Figure 8 indicates the ROI and gaze distribution of Experiment 2. There was no significant difference between Middle Skill and High Skill in the ROI of other, center, mini-map, information 1, Information 2, SD of X position, and SD of Y position in Experiment 2 (Other: p =.82; Center: p =.68; Mini-map: p =.86; Information 1: p =.32; Information 2: p =.10; SD of X position: p =.27; SD of Y position: p =.06).

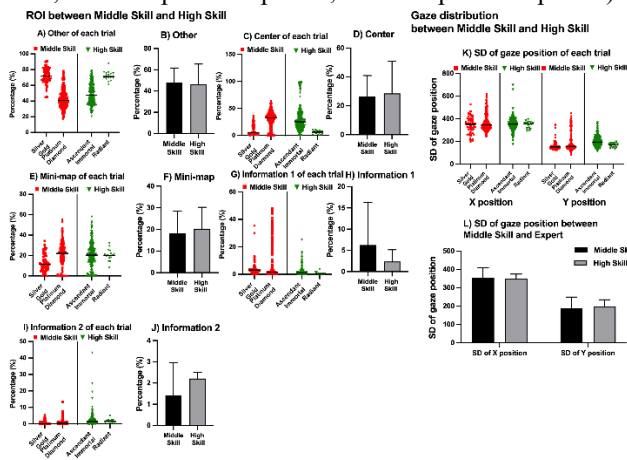

**Figure 8**. ROI percentage and gaze distribution of Experiment 2. A, C, E, G, I, and K) Each dot shows the data of each trial. B, D, F, H, J, and L) Each bar indicates the average ROI percentage and gaze distribution. Whisker shows the standard deviation of the ROI percentage and gaze distribution.

## IV. DISCUSSION

The current study aims to investigate the gaze movements of FPS game players in an actual FPS game environment (VALORANT) and a controlled environment similar to an FPS game (AimLab). In a controlled environment, high-skilled FPS game players (i.e., High Skill group) had faster reaction times than middle-skilled FPS game players (Figure 4C). This result indicates that even skilled players with similar experience (High Skill and Middle Skill) have differences in performance levels, similar to previous findings in comparing FPS gamers and non-FPS players [20]. In addition to different performance levels, our results showed that reaction time and accuracy had a positive correlation (Figure 4E). The correlation between reaction time and accuracy suggests that there may not be a speed-accuracy tradeoff effect in FPS games, which is well-known in complex motor behaviors [21, 22,23]. This is because speed and accuracy are not complementary in FPS games; FPS game players need to be fast and have accurate movement at the same time.

In the results of gaze movement data in Experiment 1, the High Skill group had a significantly narrower gaze distribution than the Middle Skilled group, and horizontal gaze distribution positively correlated with fast reaction time (Figure 5B, C, F, and G). Moreover, the distance between the center of the monitor and the center of the gaze position had a negative correlation with reaction time (Figure 6C). These results indicate that using peripheral vision is beneficial for high performance in FPS games. Peripheral vision is advantageous for reacting fast when the situation is complex, such as monitoring multiple situations [8,24]. Thus, there is a possibility that high-skilled FPS game players use their peripheral vision to react quickly. However, peripheral vision has a disadvantage in accurately checking the stimulations, probably resulting in no significant correlation between accuracy and gaze distribution during the task (Figure 5D, H, and Figure 6D).

In Experiment 2, we investigated the gaze control ability and performance level of FPS game players in actual FPS game environments. As a result of the performance level in Experiment 2, there was no significant difference between the Middle Skill and High Skill groups (Figure 7A), despite the differences in performance levels between the groups in the control environment of Experiment 1 (Figure 4C). There are two possible reasons for the lack of clear differences in the performance in Experiment 2. First, both groups had sufficient experience with the FPS game. Second, in the process of randomly assigning opponent players, it is likely that the Middle Skill group fought relatively easy opponents and the High Skill group fought relatively difficult opponent players. In Experiment 1 (i.e., the control environment), the experiment was conducted by one gamer, whereas in Experiment 2, the opponent changed, and thus, the degree of difficulty varied each match in the actual FPS game environment. Therefore, the actual game environment includes a variety of factors affecting performance, which would account for the lack of significant difference in performance between the Middle Skill and High Skill groups.

In the correlation between KDA and ROI, no significant difference was observed in each area except Information 1 (Figure 7E). Information 1 has less importance than other areas (e.g., mini-map and center area) in actual FPS games. Therefore, shifting the gaze position to view unimportant information can degrade the performance of actual FPS game players. Furthermore, the obtained results in actual FPS game environments (correlation between gaming performance and importance of information) allow us to understand the outstanding performance of FPS players with high reliability and validity, as well as sports [15]. In the result of gaze distribution and ROI in Experiment 2, there was no significant difference in gaze distribution and ROI between the High Skill and Middle Skill groups (Figure 8). Thus, it is necessary to understand the superior performance of highly skilled FPS game players under more diverse environments and skill levels.

Finally, in addition to characterizing gaze movements in FPS games, we combined an inexpensive eye tracker (Tobii 4C) with a low-cost gaze movement analyzing tool ("get_eyedata") to measure the eye movements of FPS game players in an actual FPS game environment at low cost. Compared to inexpensive eye trackers (e.g., Tobii 4C), the tools used to analyze gaze movements are expensive (e.g., Tobii Pro Lab; Funke et al., 2016) and not developed for FPS gaming. There are two advantages to combining an inexpensive eye tracker and a low-cost gaze movement analyzing tool. First, it can be used to develop a new training method for FPS game players. In general, sports, and biofeedback training methods are widely used for training [25]. Biofeedback training is a training method that uses feedback to improve the performance level and mental state [26]. According to a systematic review, the biofeedback training method can reduce injury and improve the performance of athletes [27]. Therefore, analyzing the gaze movement of FPS game players at low cost and training with appropriate feedback can help improve performance and spread the appropriate training methods in the future. Second, the current study was able to collect gaze movement data from a large number of FPS game players simultaneously. Experimenting with a combination of low-cost equipment can also help academics acquire data more easily. In the long term perceptive, it can help increase the research about FPS games and the broad esports area.

## V. Limitation

The experiment was conducted at a big esports event. Therefore, we had to acquire the data from multiple participants simultaneously and could not keep the same number of experimental trials for each participant. Moreover, because we recruited the participants from the esports event, we were unable to perform the appropriate power analysis before we started the experiment. We conducted the power analysis after the experiment. However, the power of the sample size is small (power: 0.624). Additionally, since the measurement recording frequency is low (90 Hz), it may not have been appropriate to reflect the dynamic environment of games. Additionally, further analysis is necessary to apply the developed tool to actual esports training.

## VI. Conclusion

The current study demonstrates that highly skilled FPS game players have a fast reaction time and narrow gaze distribution under a controlled environment. These results suggest that the use of peripheral vision during gameplay contributes to high performance. Furthermore, speed and accuracy were positively correlated in FPS games in a controlled environment. In the actual FPS game, we found that viewing unimportant information was negatively correlated with performance levels. Therefore, it is necessary to understand the superior performance of highly skilled FPS game players through gaze control ability in a wide range of environments. Finally, since all gaze movement data were obtained by the Python-based tool "get_eyedata" developed in the current study, the developed tool has the potential to be utilized in future esports research and training.

### Acknowledgment

We thank the GAKU Bootcamp Academy, which allowed us to conduct the research during the event. Additionally, we thank the participants who provided constructive comments and advice. The authors appreciate Donghyun Kim and Changmin Jeong, who provided advice about debugging the program and information about VALORANT.

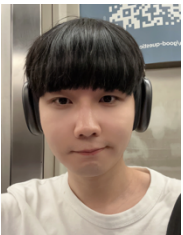
**First Author: Inhyeok Jeong** (Ph.D. student, *The University of Tokyo*) holds a BA and MA degree from Waseda University (Sport Sciences) and is currently working on esports-related research at the University of Tokyo. His research focuses on discovering the sources of superior performance in esports players, and he is interested in developing specific training methods that can be adapted to esports in the future. He is a member of the IEEE Computational Intelligence Society (CIS), Japanese Society of Physical Fitness and Sports Medicine, and The Japan Society of Coaching Studies.